\documentclass[twocolumn]{aastex62}

\usepackage{textcomp}
\usepackage{amsmath}
\usepackage{float} 
\usepackage{upgreek} 

\submitjournal{Planetary Science Journal}

\begin{document}

\title{Thermal Creep on Mars: Visualizing a Soil Layer under Tension}

\correspondingauthor{Tetyana Bila}
\email{tetyana.bila@uni-due.de}

\author{Tetyana Bila}
\affil{University of Duisburg-Essen, Faculty of Physics, Lotharstr. 1-21, 47057 Duisburg, Germany}

\author{Jonathan Kollmer}
\affil{University of Duisburg-Essen, Faculty of Physics, Lotharstr. 1-21, 47057 Duisburg, Germany}

\author{Jens Teiser}
\affil{University of Duisburg-Essen, Faculty of Physics, Lotharstr. 1-21, 47057 Duisburg, Germany}

\author{Gerhard Wurm}
\affil{University of Duisburg-Essen, Faculty of Physics, Lotharstr. 1-21, 47057 Duisburg, Germany}

\begin{abstract}
At low ambient pressure, temperature gradients in porous soil lead to a gas flow, called thermal creep. With this regard, Mars is a unique as the conditions for thermal creep to occur in natural soil only exist on this planet in the solar system.  Known as Knudsen compressor, thermal creep induces pressure variations. In the case of Mars, there might be a pressure maximum below the very top dust particle layers of the soil, which would support particle lift and might decrease threshold wind velocities necessary to trigger saltation or reduce angles of repose on certain slopes. In laboratory experiments, we applied diffusing wave spectroscopy (DWS) to trace minute motions of grains on the nm-scale in an illuminated simulated soil. This way, DWS visualizes pressure variations. We observe a minimum of motion which we attribute to the pressure maximum $\sim 2$ mm below the surface. The motion above but especially below that depth characteristically depends on the ambient pressure with a peak at an ambient pressure of about 3 mbar for our sample. This is consistent with earlier work on ejection of particle layers and is in agreement to a thermal creep origin. It underlines the supporting nature of thermal creep for particle lift which might be especially important on Mars.
\end{abstract}


\section{Introduction}

Sand and dust are mobile on the surface of Mars, readily observable from small scale dust devils over sand dune motion to global dust storms \citep{Fenton2020, Moreiras2020, Toigo2018, Lorenz2021, Heyer2020, Chojnacki2019}. This requires some lifting mechanism and gas drag related to wind and eolian transport is undoubtedly a main driver \citep{Rasmussen2015}. It has a long tradition though to argue that winds on Mars are just regularly slightly too weak to account for particle lift \citep{Greeley1980, Kok2012}.

There are ways where rather low shear stress might suffice. E.g., the continuation threshold for saltation is way lower than the initial threshold \citep{Kok2010b}, aggregates are easier to be picked up \citep{Merrison2007}, settling at low gravity might make the soil more prone to wind drag \citep{Musiolik2018}, sporadic motion by gusts might also contribute \citep{Swann2020}, and electrostatic forces might play a role as well \citep{Esposito2016, Kruss2021}. 
So, it might be debatable if the wind on Mars has a general problem to initiate particle motion but, in any case, if at its best the wind speed is always only close to the entrainment threshold velocity, then every other effect might play a very important role in providing conditions in favor of particle mobility \citep{Neakrase2016}.

Among the supporting mechanisms for particle lift or downhill flow might also be sub-soil processes. Resurfacing water in selected locations has been discussed as a sediment transport mechanism \citep{Raack2017}, also pressure excursions coming along with the vortex of a dust devil might provide lift to surface material \citep{Balme2006, Koester2017, Bila2020}.  Especially in this latter case, the pressure within the soil is higher than above the soil. 

The ambient pressure range of mbar allows one more lifting mechanism on Mars which also leads to a subsurface overpressure, eventually. This is based on a Knudsen compressor induced by thermal creep within the soil as the surface is insolated. We explain this in detail in the following section. The effect can be large but occurs on a small scale and therefore is intrinsically difficult to be measured directly. Therefore, a visualization is the focus of our work here.

\section{Thermal creep induced sub soil overpressure}

\begin{figure}[p]
 \centering 
\includegraphics[width=0.823\columnwidth]{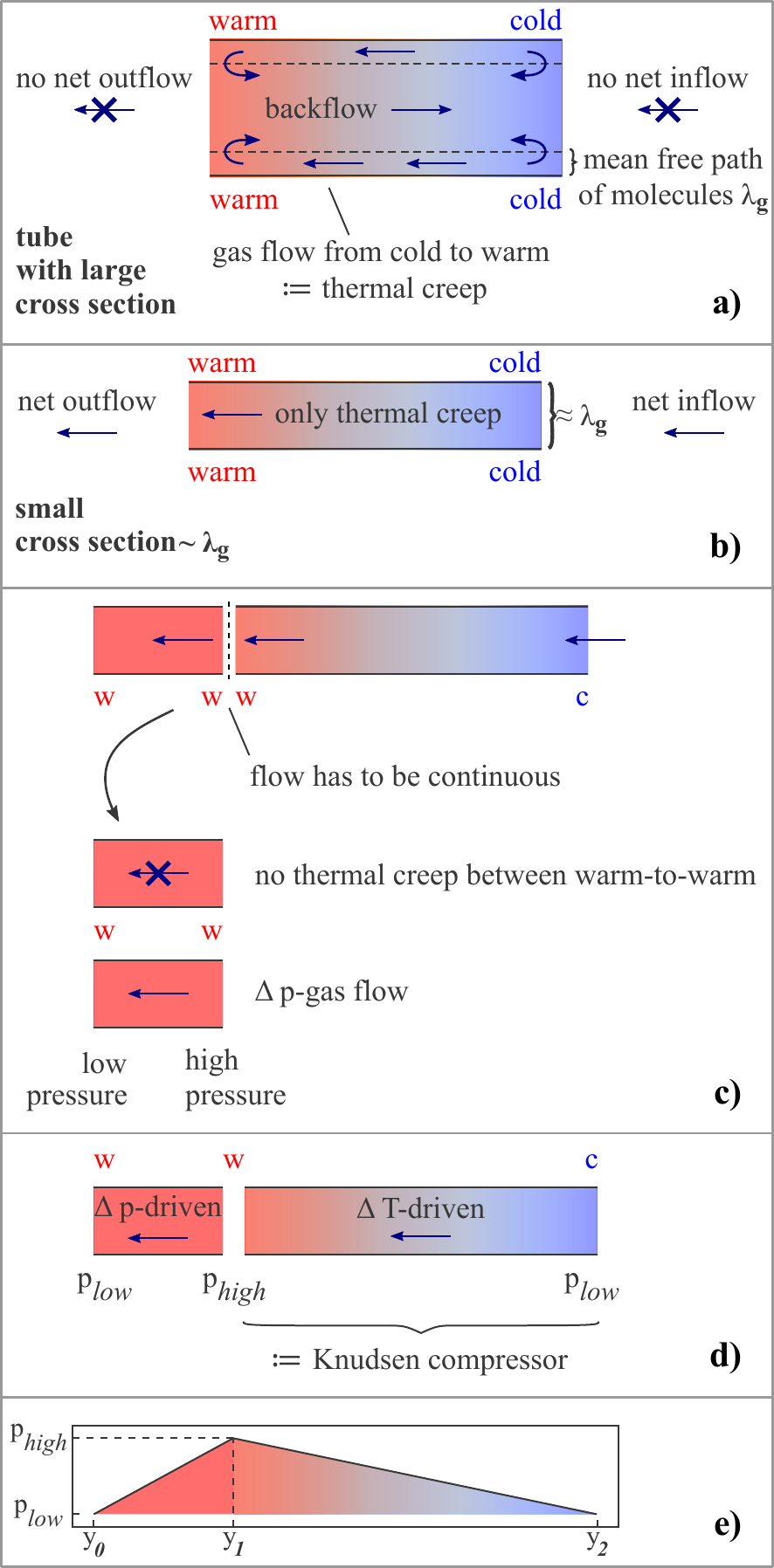}

 \caption{ \small \label{sketchi} The basic principle of thermal creep and Knudsen compressor. a) Gas flow in an open tube of the large cross section with a temperature gradient between the tube ends. In a tube with a large cross section, there are two opposing gas flows: a gas flow from the colder end to the warmer end in a thin layer on the order of the mean free path of molecules $\lambda_g$, called thermal creep, and a backflow from the warmer to the colder end. b) In a tube with a cross section on the order of $\lambda_g$, the thermal creep gas flow dominates. c-d) A tube with a cross section on the order of $\lambda_g$ with a temperature gradient between its ends. On the warmer side the tube is followed by a smaller tube of the same cross section having a constant temperature. The thermal creep gas flow cannot proceed into the smaller tube and the pressure increases at the junction. This leads to a pressure-driven gas flow through the smaller tube. The pressure adjusts itself until the incoming temperature gradient-driven gas flow equals the pressure-driven gas flow. e) In-depth basic pressure distribution in a dust bed heated from above. $y_0$ denotes the surface of a particle bed. In the near-surface layer from $y_0$ to $y_1$ the temperature is constant. Below $y_1$ down to $y_2$ the temperature gradient is present.}
\end{figure}

The underlying physical principle of thermal creep and Knudsen compressors are not widely spread and to some degree counter intuitive, so we review the basics here along the lines of fig. \ref{sketchi}. Part "a" shows the general gas flow situation in a single open tube with a temperature gradient along its wall. In a thin layer on the order of the mean free path of the molecules $\lambda_g$, gas molecules then get a net component of motion in the direction from cold to warm. This is owed to the non-thermal equilibrium situation and is called thermal creep. It is just that at a given constant pressure, on the cold side the rate of molecules traveling to the warm side is higher than the other way round. As one usually recalls heating a gas reservoir with expansion or pressure increase and therefore a motion from warm to cold, this flow is not very intuitive. In large tubes, it also essentially goes unnoticed without larger effects outside the tube as there is ample space in the center of the tube to lead to a backflow without much pressure difference needed.

This changes for small tubes with an overall diameter on the order of the mean free path as pictured in fig. \ref{sketchi} b.
Thermal creep now dominates over the total cross section. There can certainly still be a backflow but now this would need to be driven by a significant pressure gradient. Therefore, if, e.g., the end of the tubes would be closed, this would lead to a pressure increase on the warm side until thermal creep flow and pressure-driven backflow would balance. This would then be called a Knudsen compressor (s. fig. \ref{sketchi} d) according to Knudsen who studied the pressure conditions at the end of small tubes in the early twentieth century \citep{Knudsen1910}. At perfect conditions, he could realize situations where the ratio between pressures at the warm to the cold side was on the order of factor 10.

So at its extremes, thermal creep in a tube can either lead to a gas flow of maximum rate, if the ends of the tube are open, or thermal creep leads to a maximum pressure increase, if the flow cannot proceed, i.e. if it is blocked by walls or if there are adjacent volumes but if these are negligible in sizes. The situation relevant here is somewhere in between and pictured in fig. \ref{sketchi} c. If we just add a (small) tube on the warm side which is at a constant temperature, then there is no thermal creep that can transport the incoming flow further on, so the pressure increases and the flow speed decreases until the pressure gradient along the small tube transports the flow onward and the temperature-driven thermal creep flow equals the pressure-driven gas flow.

As outlined and entering in detail, thermal creep effects depend on the ratio $Kn$ between the mean free path of the gas molecules $\lambda_g$ and tube diameter $d$. This is known as Knudsen number. If the Knudsen number is small, tubes are relatively large and no pressure can build up. If the Knudsen number is high, the tube is very small and the flow rate eventually gets too small to sustain pressure differences. Therefore, thermal creep effects are most pronounced under conditions where the Knudsen number is around $Kn \sim 1$.

\begin{figure}[htb]
 \centering
\includegraphics[width=0.9\columnwidth]{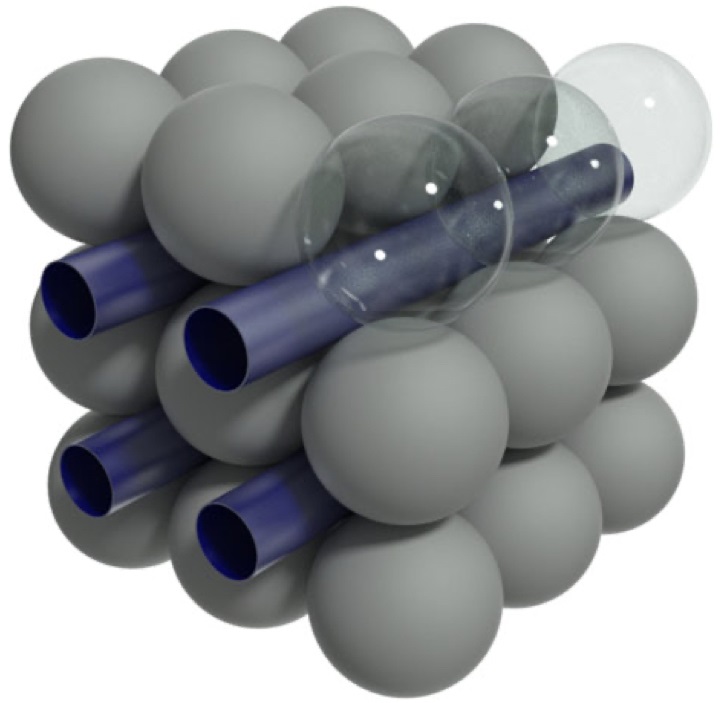}

 \caption{\small \label{flow}Model of a granular medium as a collection of pore-size tubes (from \citet{Steinpilz2017}).}

\end{figure}

To make the connection to Martian soil, the question is how the tube picture can be applied
to a porous granular bed. While the idea would be the same, namely that thermal creep transports gas through the pores, this situation is more complex. There are pores of varying size. So for one thing, the Knudsen number is not well defined \textit{a priori}.
A simple analogy is shown in fig. \ref{flow} from \citet{Steinpilz2017}, where the granular bed is idealized by a regular cubic structure.  \citet{Koester2017b} studied various real particle beds experimentally and showed that indeed a granular medium can very well be described by a set of tubes where the diameter of the tube essentially equals the average particle (pore) diameter, which would be slightly larger than the sketch suggests. 

Finally, as with the two tubes in fig. \ref{sketchi} c, we can  have different parts of a particle bed or soil. Thermal creep can pump gas without the need for a pressure difference in a region with a temperature gradient. If this connects i.e. to a top layer at a constant temperature, pressure builds up until gas is driven further in the usual way by a pressure difference which is known as Darcy flow. More detailed equations for calculating thermal creep for granular matter can be found in \citet{Koester2017b}. The pressure gradient-driven top layer under tension has a certain thickness, which we visualize below as the overpressure will always lead to at least minute particle motions upwards and downwards and there will be a minimum of particle motion close to the pressure maximum.

There have been observations of particle motions before but at such large levels of incident light flux that the pressure gradients were large enough to eject grains directly against gravity and cohesion. First experiments on particle lifting at low pressure when illuminated were carried out by \citet{Wurm2006} but among the possible drivers, all related to non-equilibrium low pressure gas physics, thermal creep generated overpressure was only recognized by \citet{debeule2014}, supported by a number of works in between  \citep{Wurm2008, Kelling2009, Kelling2011, Kocifaj2010, Kocifaj2011}.

\citet{Kelling2011} find aggregates on the order of 100 $\rm \mu m$ being ejected in experiments. Numerical simulations in \citet{Kocifaj2011} show that the temperature changes from flat to strongly decreasing at a few mm depth. The layer under tension was probed by laboratory measurements by \citet{deBeule2015} based on the thickness of an ejected layer to be up to a few 100 $\rm \mu m$.

This overpressure might not be strong enough on its own under Martian sunlight but can reduce the threshold shear velocity for particle lift by about 20 \% \citep{Kuepper2015}.
The effect can be amplified under certain conditions of illumination, i.e. at a shadow boundary \citep{Kuepper2016}. This might help explaining Recurring Slope Lineae on Mars \citep{Schmidt2017}.

In any case, the top layer that is active is rather thin.
While this is consistent with temperature profiles modeled, this is only on the order of the grain size or slightly larger. Therefore, the pressure profile cannot be probed directly.

In the spirit of evaluating different methods for tracing the pressure or at least the maximum pressure line, we employed a new method here, called diffusing wave spectroscopy and we present first measurements which qualify this technique as suitable visualization tool for thermal creep pressure variations.

\begin{figure}[p]

\includegraphics[width=\columnwidth]{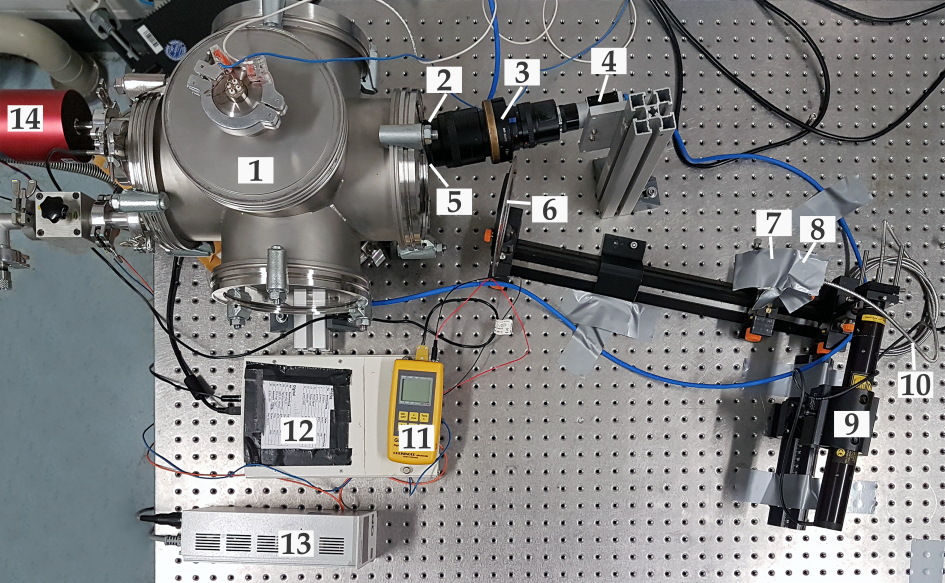}

\vspace{0.25cm}

\includegraphics[width=0.964\columnwidth]{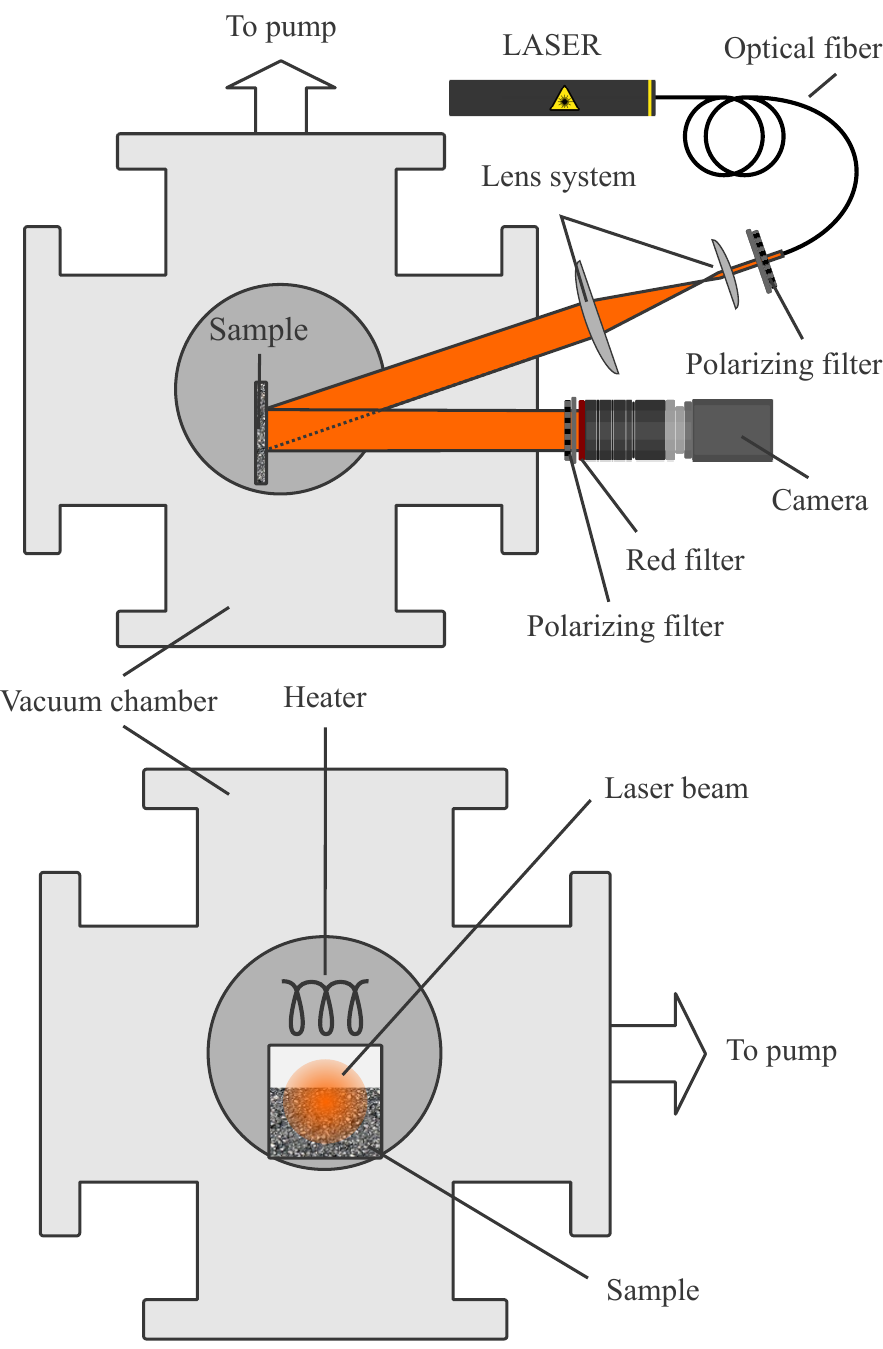}

 \caption{ \small \label{fig:AufbauOben} Top: The experimental setup used. 1) Vacuum chamber, 2)  polarizing filter, 3) camera lens, 4) camera, 5) red filter, 6) lens with f=300 mm, 7) lens with f=10.8 mm, 8) polarizing filter, 9) laser, 10)  optical fiber, 11) digital thermometer, 12) analog-to-digital converter, 13) pressure display unit and 14) vacuum gauge. Middle and bottom figures show the sketch of the setup in top view and side view respectively.  A granular soil simulant sample within a vessel is placed in a vacuum chamber. Heating from the top (IR radiation) simulates insolation. The speckle pattern of a laser backscattered from the sample from the side is imaged by a camera.
}
\end{figure}

\begin{figure}[htb]
\includegraphics[width=\columnwidth]{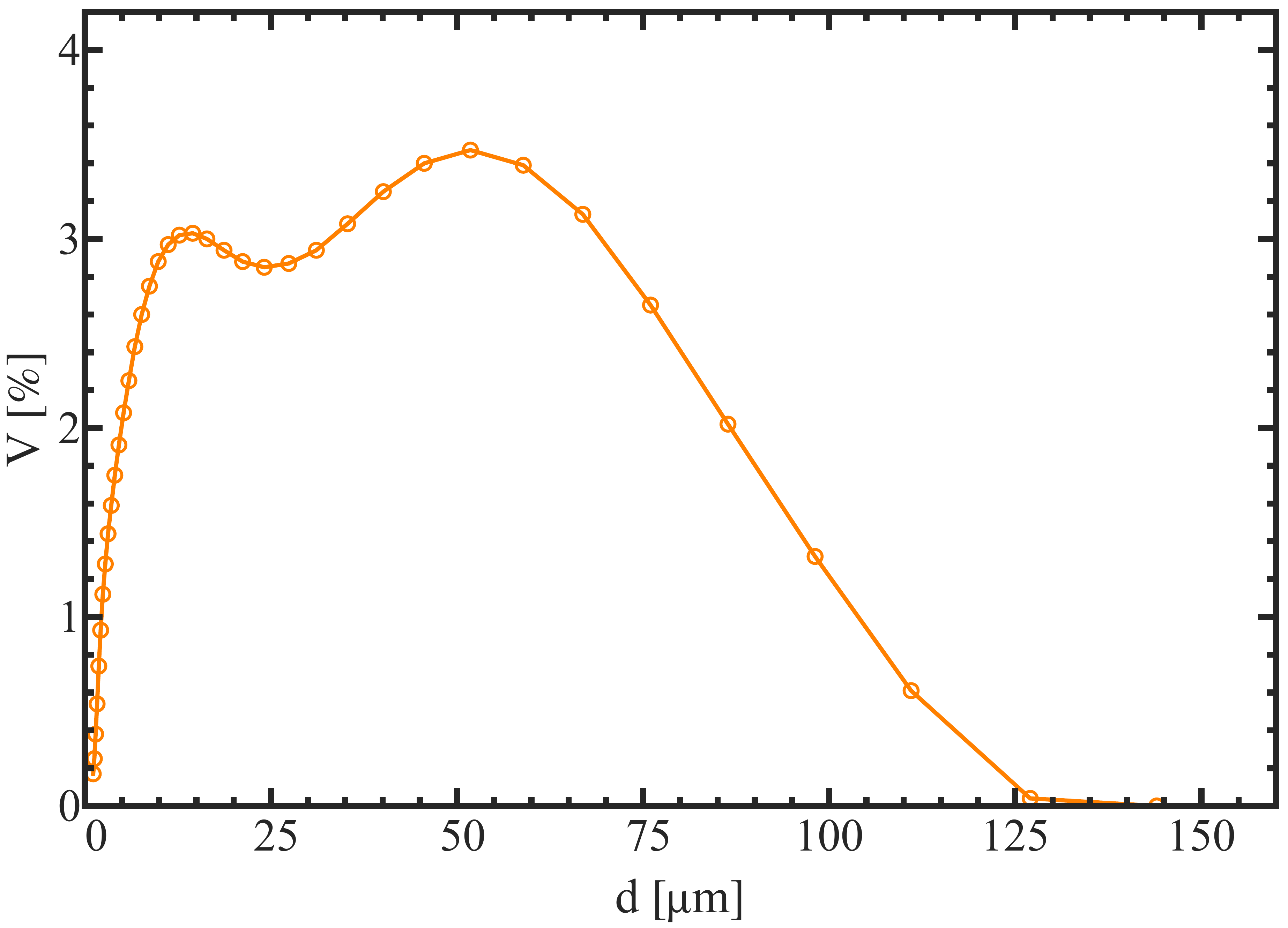}
 \caption{ \small \label{fig:Groessenverteilung} Volume size distribution of the basalt sample used. From this distribution the mean particle diameter is calculated to $51 \pm 28 \, \mathrm{\upmu m}$.
}

\end{figure}

\pagebreak

\section{Experiment\label{sec:experiment}}

\autoref{fig:AufbauOben} middle and bottom presents a sketch of the experimental setup. A vessel filled with sample soil material is placed inside a vacuum chamber. 
The volume of the vessel holding the sample is $(9.5 \times 9 \times 1) \, \mathrm{cm^3}$, where the sample takes a volume of $(3.8 \times 9 \times 1) \, \mathrm{cm^3}$. One side of the vessel is transparent with a 2 mm flat glass plate. 

Heating of the surface by infrared radiation is induced by a heated wire placed about $3 \, \mathrm{cm}$ above the sample surface.
We did not quantify the radiative flux here as our focus is on the DWS technique and what it might provide as information on particle motion.
As sample we use basalt with a broad size distribution (\autoref{fig:Groessenverteilung}) and a mean diameter of $51 \pm 28 \, \mathrm{\upmu m}$. The bulk density of the sample is about $1.7 \, \mathrm{g/cm^3}$, which results in a porosity of about 0.44.

A red (633 nm), 2 mW beam shaped ($\sim$ 3 cm diameter) laser illuminates a large part of the soil from the side, especially including the surface region. 
The scattered light is recorded by a video camera. The optical pathways of laser and camera hold crossed linear polarizers to exclude geometrically reflected light but limit the detection to light scattered by the sample which changes the polarization of the incoming radiation. The camera is focused on the front glass plate plane and has
a resolution of $10 \, \mathrm{\upmu m/pixel}$. Due to constructive and destructive interference it records a speckle image that changes with slight movements of the sample particles.

The experiment was carried out at 8 different pressures $1.9 \pm  0.1$ mbar, $2.5 \pm 0.1$ mbar, $3.6 \pm 0.1$ mbar, $5.2 \pm 0.1$ mbar, $8.2 \pm 0.1$ mbar, $148.9 \pm 0.4$ mbar, $293.9 \pm 0.6$ mbar and $780 \pm 2$ mbar. The initial pump-down took tens of hours (a weekend) in order to assure that most of the water within the pore space is evaporated and not influencing the measurements. Final adjustments were made before each measurement using a precision valve. The temperature before the heater was switched on was $296.1 \pm 0.5$ K for all measurements. The maximum temperature at the end of the measurement depends on the pressure. \autoref{fig:tiefendruck} (bottom) shows the pressure dependence of the absolute temperature increase ($\Delta T$) of the top soil layers.

\begin{figure}[tb]
\centering
\includegraphics[width=0.98\linewidth]{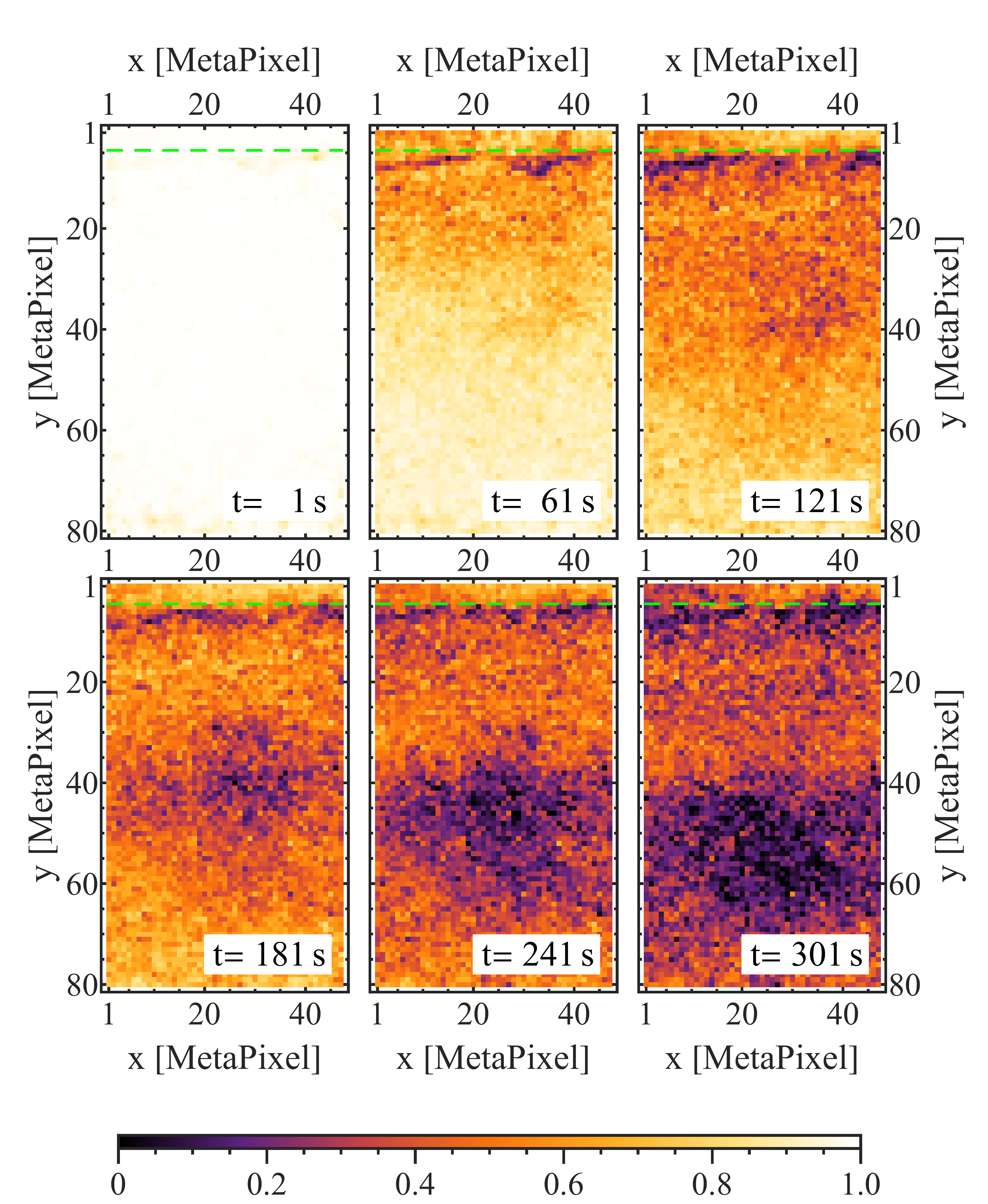}
 \caption{ \small \label{fig:CORRt} Example of a time evolution of the correlation {for the measurement at $5.2 \, \mathrm{mbar}$}; times after the heater is switched on are imprinted in each image. The dashed green line indicates the position of the sample surface. $y$ denotes downward {direction} and $x$ is the horizontal. The size of one metapixel is 20 pixel.}
\end{figure}

\section{Diffusing wave spectroscopy}

A speckle pattern is the essential part of diffusing wave spectroscopy, which is a technique which visualizes very small displacements in granular materials \citep{Crassous2007,Erpelding_2010, Amon2017, brown1993}. In the used backscattering geometry, the laser beam illuminates the granular bed from the side. The light is then entering the granular medium, is scattered several times on various grains and part of the radiation leaves the sample again on the same side. As mentioned, due to the coherent nature of the light source and interference of the various scattered beams, a focused image results in a speckle image with bright spots for constructive interference and dark ones if pathways result in destructive intereference.

The scattering medium, in our case the soil simulant, is characterized by its transport mean free path $\lambda$. On their way through the sample, the photons scan a volume on average in the order of $\lambda^3$ \citep{Erpelding2008}. Hence, the scattering process limits the spatial resolution of the method to the transport mean free path. Several pixels of an image are grouped into a \emph{metapixel}. In our case, the speckles have a size of about $2 $ pixel $(20 \rm \upmu m)$ and we take 20 by 20 pixels as metapixel.
$\langle I \rangle$ is the average of the brightness $I$ over all pixels inside a metapixel.

If grains shift on the size scale of the wavelength, the speckle pattern changes. Therefore, areas where displacements take place can be visualised in a correlation plot comparing corresponding metapixels from images taken at different times $t$. The intensity correlation function $G_l$ is calculated for the corresponding $l$-th metapixel between the initial frame at $t=0\,$s and one of the following frames as 

\begin{equation}
\label{eq:Gi}
G_l = \frac{ \langle I_\textit{0l} \cdot I_\textit{tl} \rangle - \langle I_\textit{0l} \rangle \langle I_\textit{tl} \rangle}{ \sqrt{ \langle I_\textit{0l}^2 \rangle - \langle I_\textit{0l} \rangle^2 } \cdot \sqrt{ \langle I_\textit{tl}^2 \rangle - \langle I_\textit{tl} \rangle^2 }}.
\end{equation}

In eq. \ref{eq:Gi}, each metapixel $l$ of a frame is represented by its brightness $I_\textit{tl}$ where t denotes the frame at time $t$. 
$\langle I_\textit{0l} \cdot I_\textit{tl} \rangle$ denotes the average of the element wise multiplication of respective pixels within the two metapixels.

\begin{figure}[t]
\centering
\includegraphics[width=0.88\columnwidth]{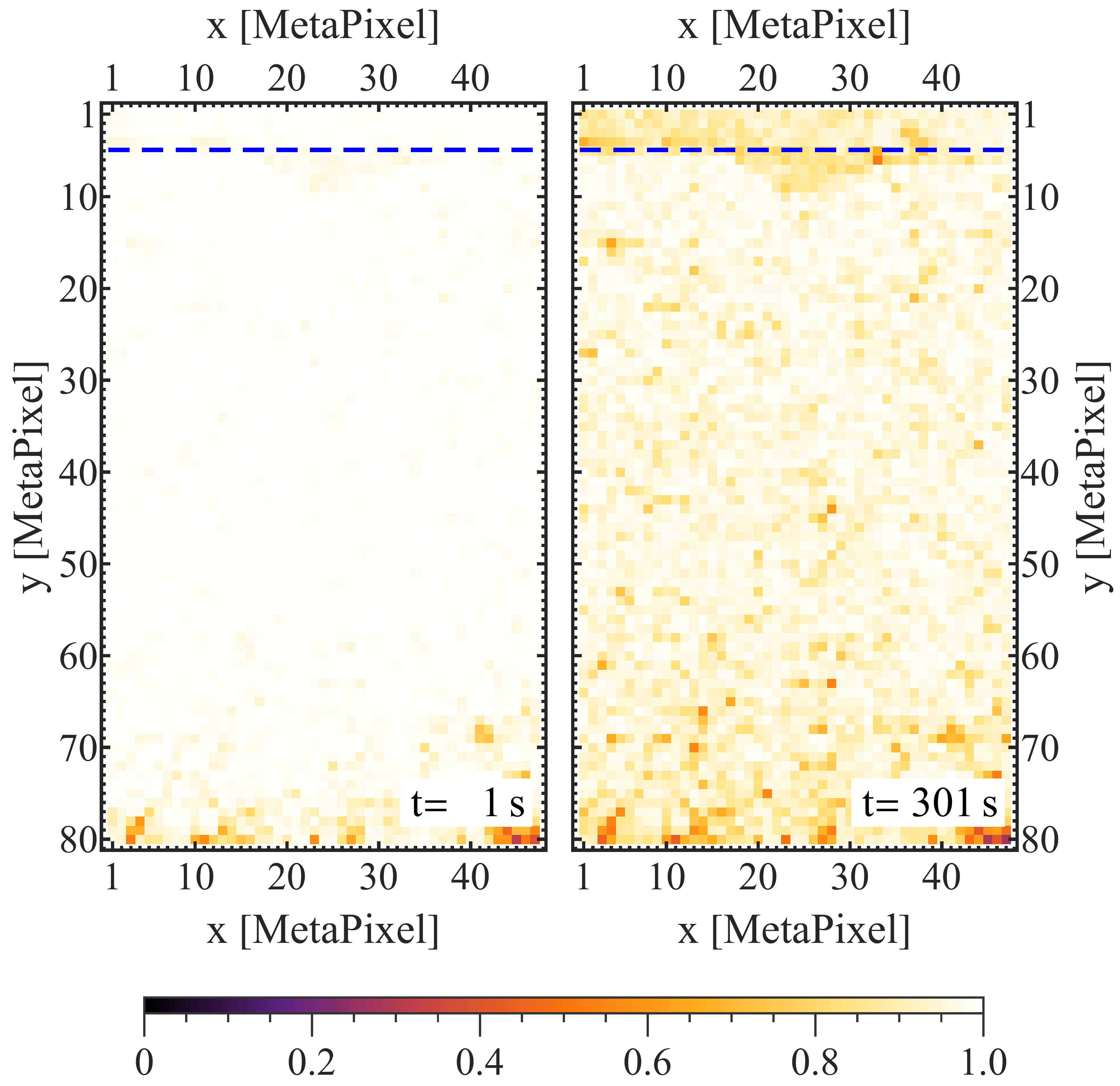}
 \caption{ \small \label{fig:CORRohneHeizen2} Same as fig. \ref{fig:CORRt} but without heating at $\sim 1190$ mbar ambient pressure.}
\end{figure}

At $t=0\,$s the sample is undeformed. The deformation develops with time and therefore can be seen as a decrease of the correlation value. A correlation value close to 1 indicates that no or very weak displacement of scattering particles has taken place.
The correlation values can be plotted spatially resolved, building a correlation map where each correlation value is a measure for the deformation occurring in a cell of $\lambda^3$-size.

Fig. \ref{fig:CORRt} shows an example of the correlation plot for the heated sample in the illuminated part of the vessel over time. Fig. \ref{fig:CORRohneHeizen2} shows the same as fig. \ref{fig:CORRt}, but for the sample at room temperature without heating.

\section{Results and discussion}

\begin{figure}[t]
 \centering 
\includegraphics[width=\columnwidth]{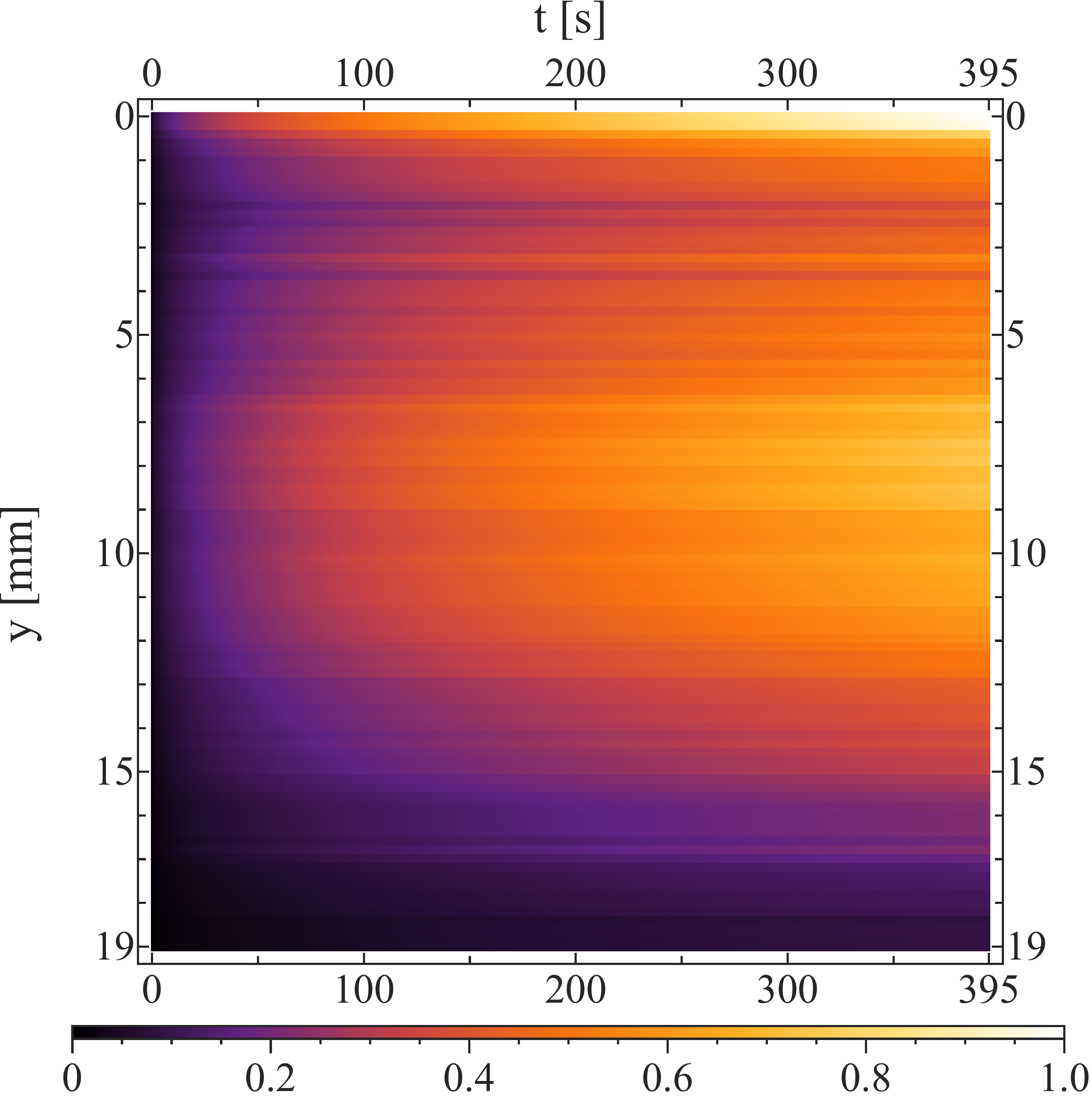}
 \caption{ \small Strength of the motion of grains with depth {y} over time at {5.2 mbar} ambient pressure. Dark color marks little motion (reversed color from correlation plot fig. \ref{fig:CORRt}).}
\label{fig:daten8mbar}
\end{figure}

To characterize the depth dependence of the particle motion in more detail, we average the data horizontally.
Fig. \ref{fig:daten8mbar} then shows the time evolution of the depth profile in absolute spatial units now. We reversed the colors here for visibility, i.e. bright colors refer to large motions.
Especially deeper layers down to 10 mm loose correlations over time, i.e. show large particle motion on timescales of minutes. This is consistent with heat conduction timescales. We did not follow the evolution longer than 400 s here. While heat is transported further downwards, the correlations do no longer change in the top layers, eventually.

\begin{figure}[p]
 \centering 
{\includegraphics[width=0.75\linewidth]{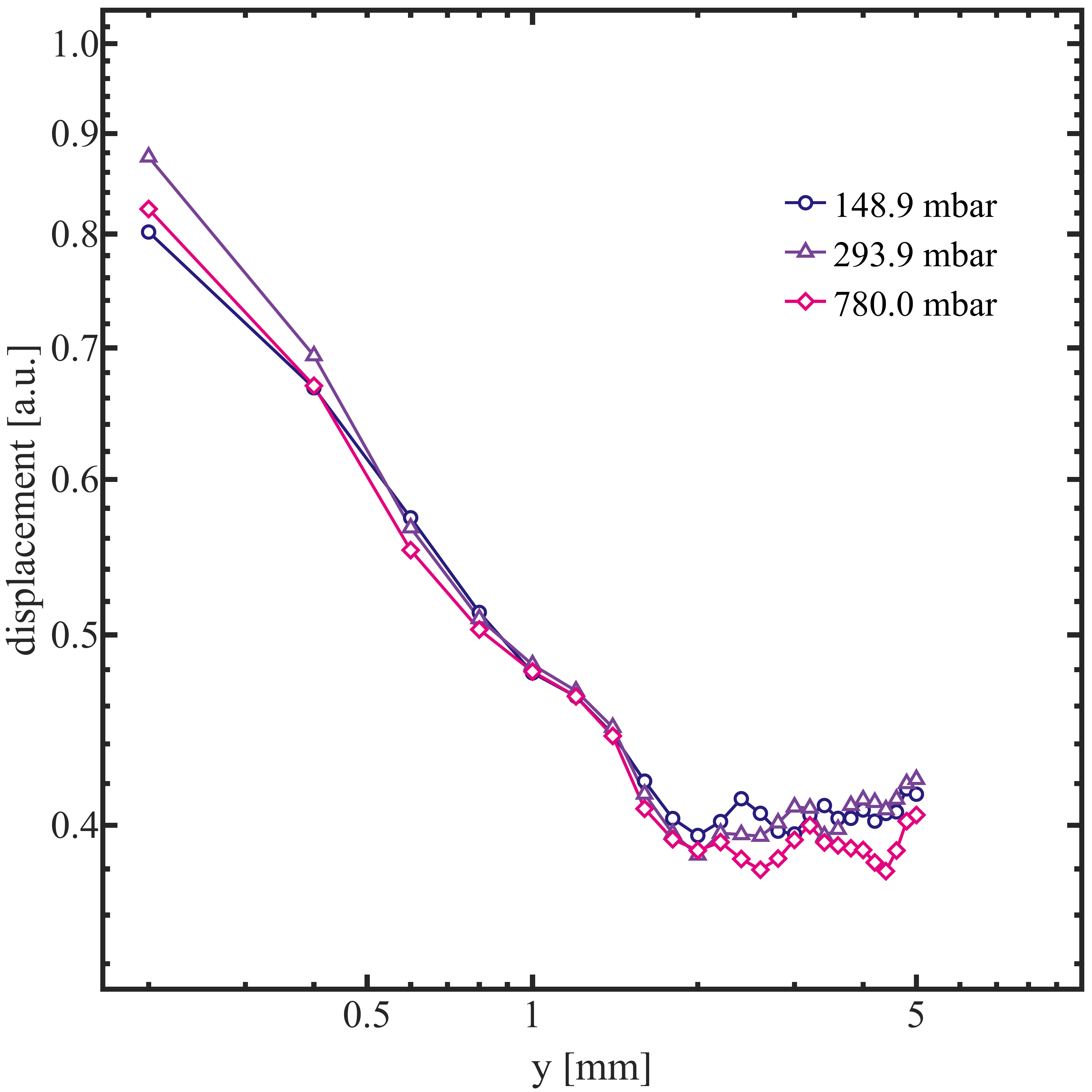}}
{\includegraphics[width=0.75\linewidth]{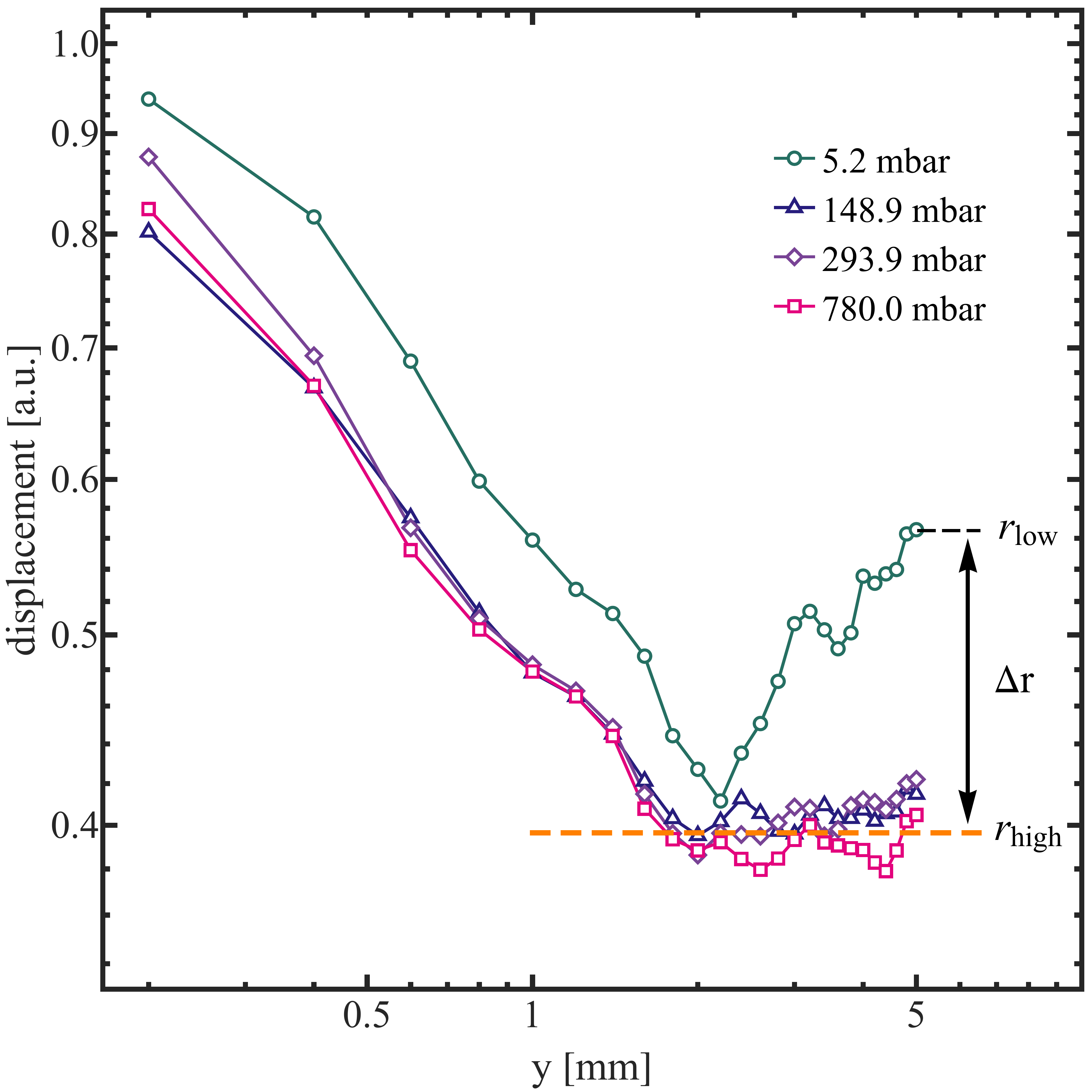}}
{\includegraphics[width=0.75\linewidth]{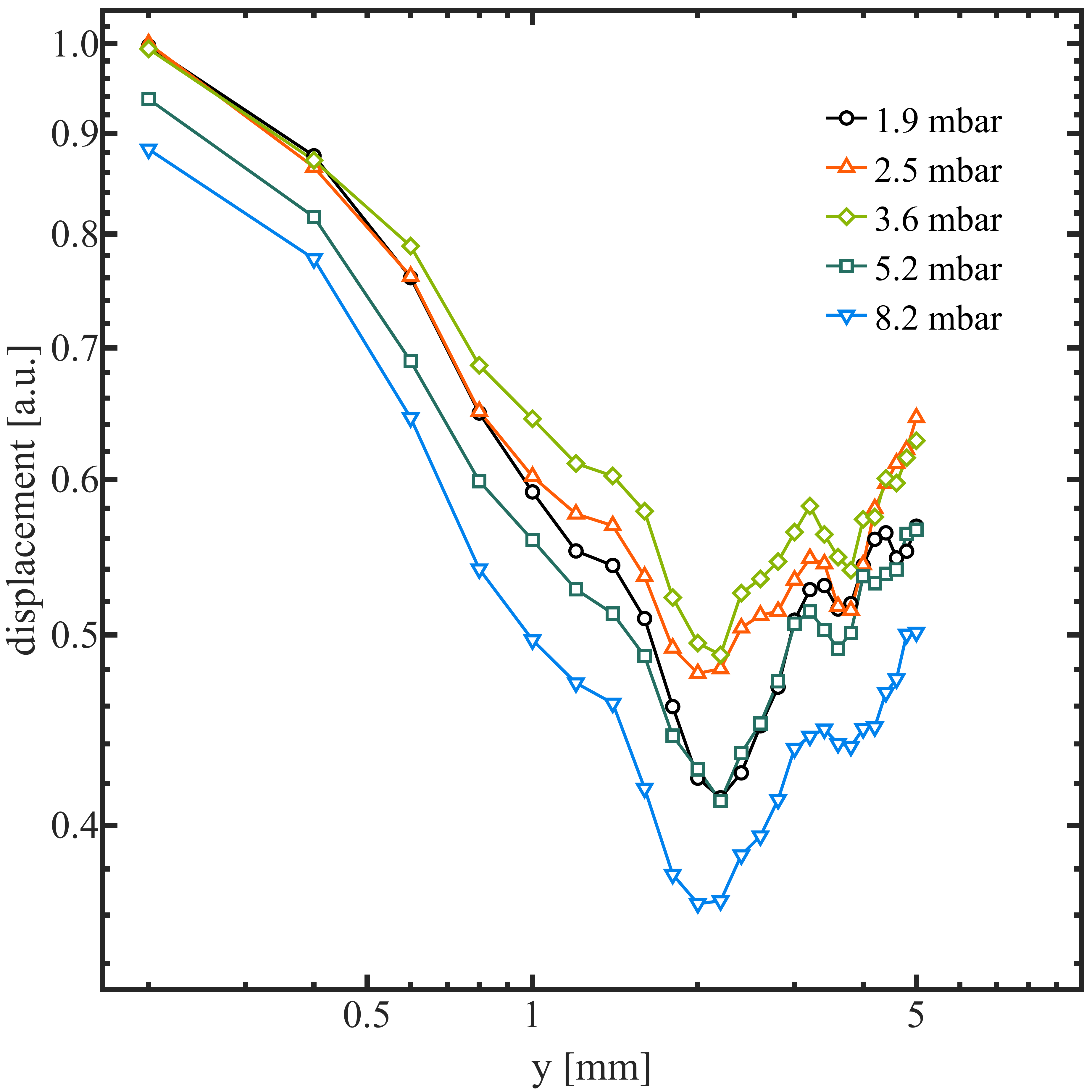}}
 \caption{ \small Motion depth profiles at the latest time (395 s); top: high pressure data with clear split in two parts, a downward decreasing and then nearly constant part; middle: overplotted example of low pressure data with an additional increase at larger depths. The orange dashed line is a fit with $r_{high}=const.$ on the flat parts of all three of the higher pressure depth profiles. $r_{low}$ is the value of the depth profile of lowest pressure at $y=5 \,$mm. $\Delta r$ is the difference between $r_{low}$ and $r_{high}$; bottom: all low pressure data. 
}
\label{fig:displacementVonP}
\end{figure}

To discuss the depth profiles we therefore then choose the profile of the latest time at 395 s for each of the 8 pressures sampled. These are shown in fig. \ref{fig:displacementVonP}.  

To show the characteristics of the depth profiles we divide the data into different pressure ranges. Fig. \ref{fig:displacementVonP} top shows the depth profile for the three highest pressures from about 149 mbar to 780 mbar. No thermal creep should be present at this high pressures and, indeed, the curves are all essentially the same. Here we focus on the depth down to $5\,$mm.
These motion depth profiles at higher pressure can clearly be split into two parts. Down to 2 mm, the motion decreases roughly like a power law (linear in log-log plot). Below 2 mm the motion profile is flat. As thermal creep is not acting, the motion close to the surface has to have a different origin. We cannot pin down the reason but in general, the topmost grains are bound the least, are not compressed by the grains close by and therefore are the most mobile. So they are moved the easiest by any disturbance. The further down, the more confined are the grains until disturbances are too faint to result in any motion. Disturbances might be simple gas flow or expansion as the sample heats up but, again, we do not know the exact reason. In any case, these profiles provide a firm base for comparison to the lower pressure range. 

To make the difference clear, we now overplot one example of a depth profile for the lower pressure range (5.2 mbar) in fig. \ref{fig:displacementVonP} middle where we expect thermal creep to act. For completeness, we plot all low pressure data in fig. \ref{fig:displacementVonP} bottom. In comparison to the higher pressure range, the low pressure data have a simple but very distinguished shape with 3 points to note. 
\begin{itemize}
    \item The topmost decrease of motion is still a power law and essentially one with the same power as for higher pressures but the absolute value is shifted upwards. This implies more motion in the top layer. This is consistent with a subsurface overpressure with a pressure gradient above that moves all grains up to the very top somewhat more.
    \item There is a clear minimum for all low pressure data at about 2 mm. The position of this minimum does not change significantly.
    \item Finally, the motion now increases again deeper within the particle bed. This is also consistent with a subsurface overpressure that works in both directions upwards but also downwards.
\end{itemize}
The motion of grains requires forces acting on them. These can be provided by the sub-soil pressure gradients induced by thermal creep. As indicated in fig. \ref{sketchi} e, there is a pressure maximum close to the surface, as the upward thermal creep gas flow can only be maintained by a pressure-driven gas flow in the upper layers with little temperature gradient. Further down below the surface, along the flat temperature distribution in the near-surface layer, there will be another transition to a flat temperature distribution, which might result in pressure variations, though this is not shown in fig. \ref{sketchi} d which was focused on the top layer.

It seems reasonable to assume that these pressure variations drive the motion of grains seen in fig. \ref{fig:daten8mbar}. 
It has to be noted though that the location of particle motion and the location of gas pressure extrema might not necessarily coincide 1 to 1 \textit{a priori}. The gas pressure gradient implies certain local forces on the grains but forces on grains will in addition be commuted to the other grains by force chains.
In other words, the minimum of particle motion traces an equilibrium spot where upward and downward directed forces would balance but even a bulk motion, i.e. large scale lift or compression would be seen as motion in DWS as it always comes with small variations. Large scale motions are not present in the data though. The sample further below the surface only moves significantly once the temperature rises in that region as the heat is conducted downwards. Any motion further down in the particle bed occurring then is not significantly influencing the layers above, so pressure gradients and particle motion can be considered to act only locally.

\begin{figure}[htb]
 \raggedright
{\includegraphics[width=0.97\linewidth]{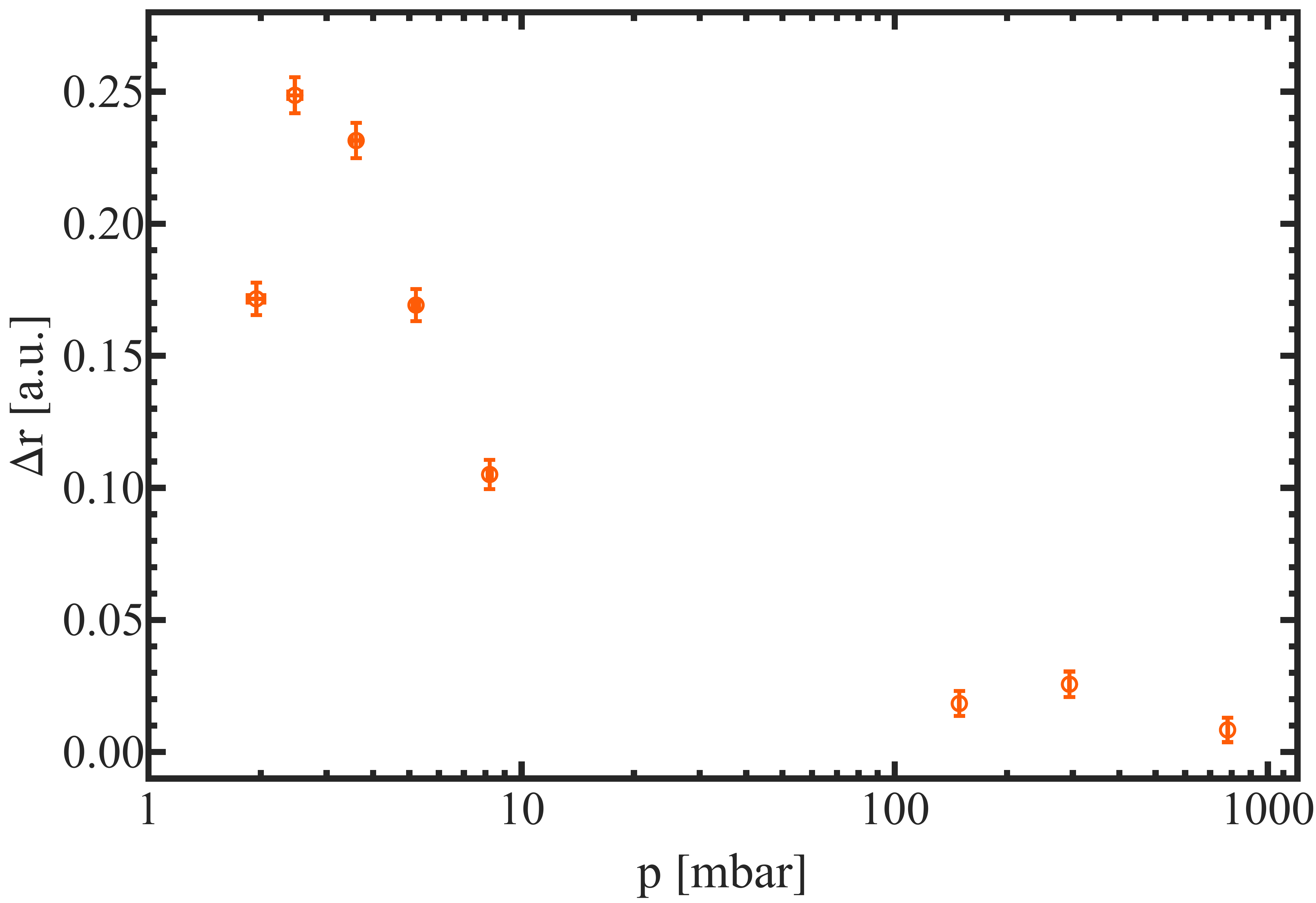}}
{\includegraphics[width=0.97\linewidth]{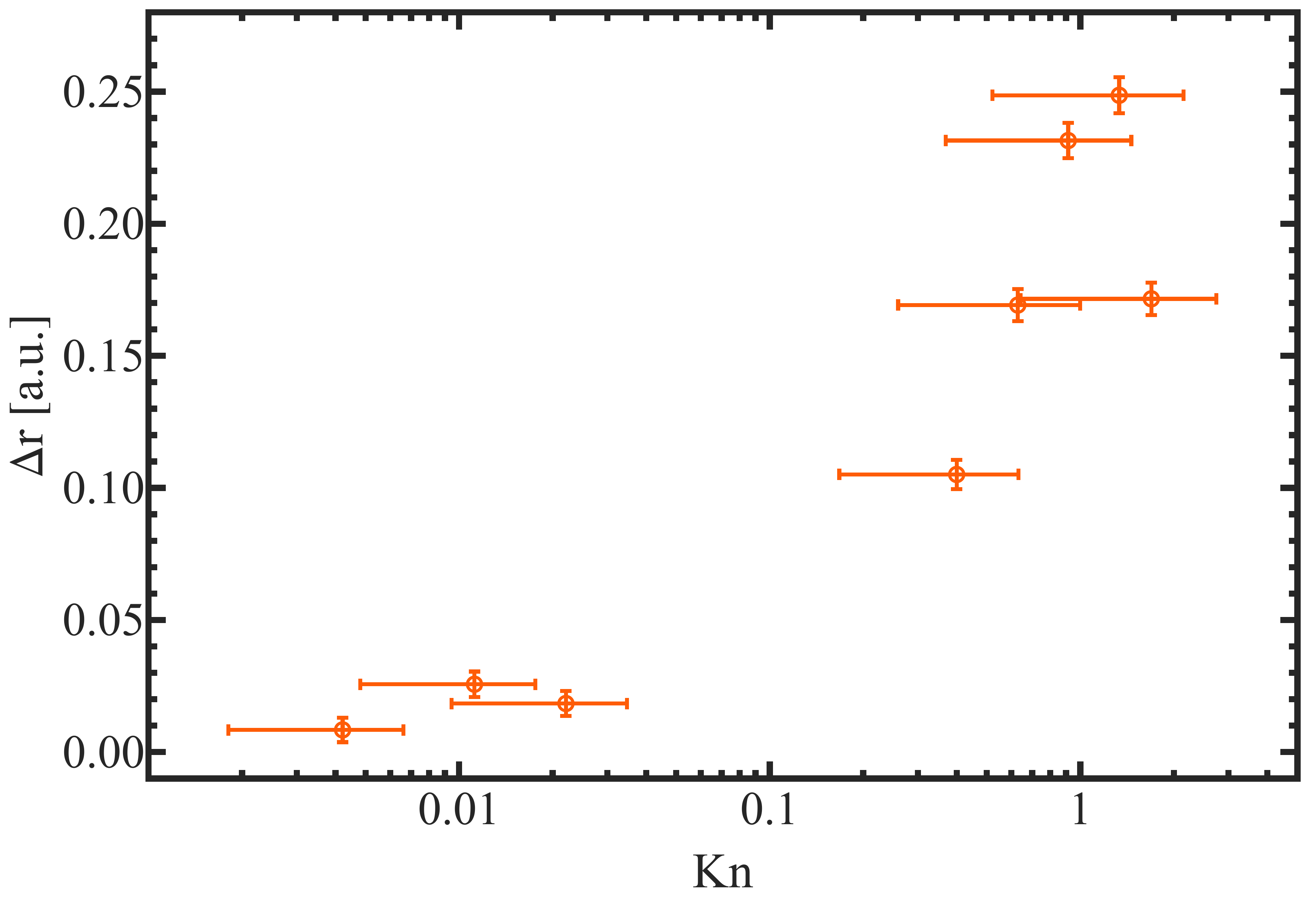}}
{\includegraphics[width=0.97\linewidth]{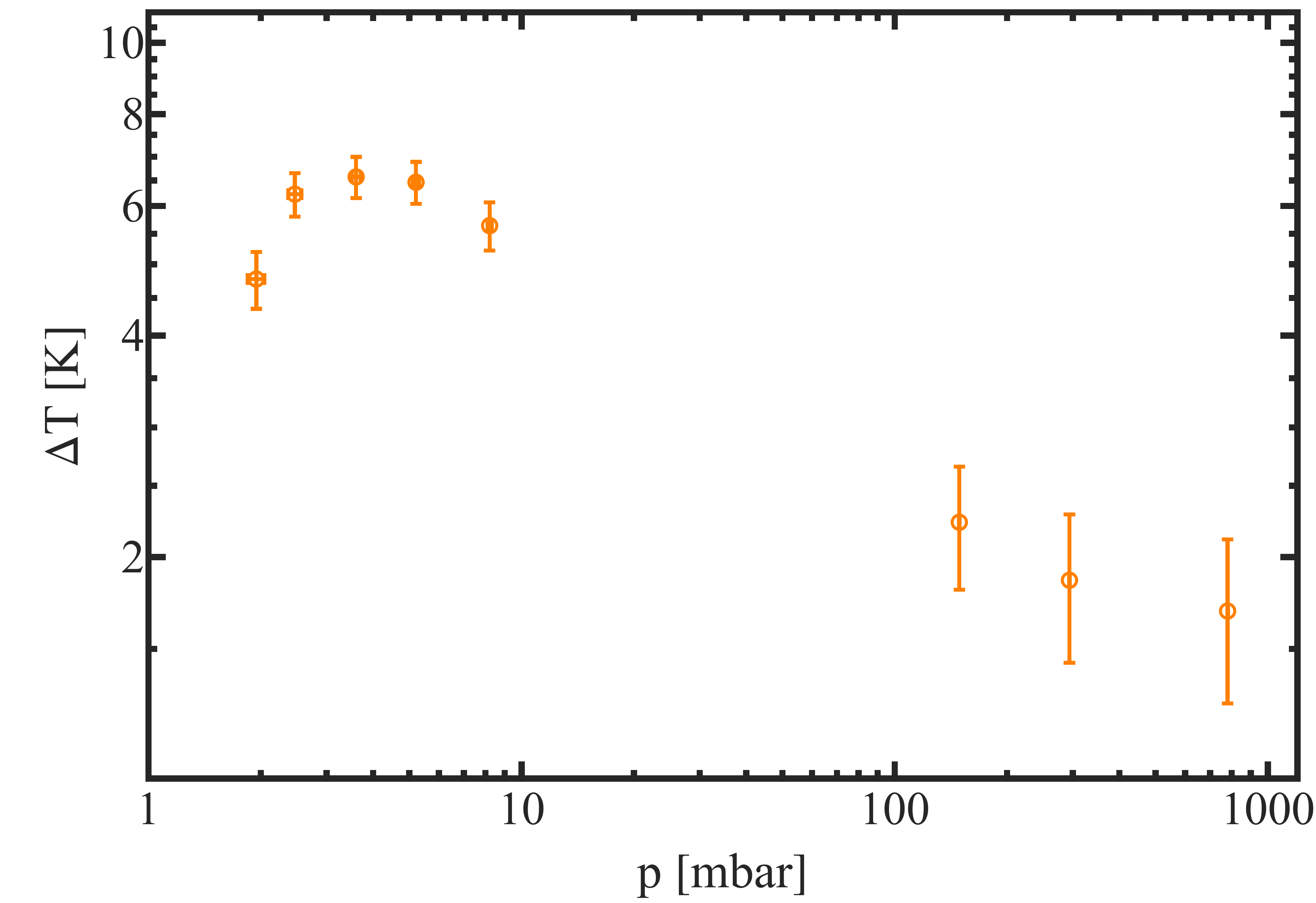}}
 \caption{ \small Top: pressure dependence of $\Delta r$; Center: dependence of $\Delta r$  on Knudsen number; Bottom: Temperature difference at the end of the measurement over the ambient pressure at the beginning of the measurement. 
}
\label{fig:tiefendruck}
\end{figure}

The most pronounced differences with ambient pressure are the strengths of the motion increase below the surface.
We therefore studied the differences $\Delta r= r_{low}- r_{high}$ over pressure and, indeed, we observe a systematic variation as shown in fig. \ref{fig:tiefendruck} (top). The effect is strongest at an ambient pressure of about 3 mbar rapidly decreasing to the offset value as the pressure is increased by a few mbar. In terms of the Knudsen number $\rm{Kn} = \lambda_g / d$ with the mean free path of the gas molecules $\lambda_g$ and an efficient pore diameter of $d = 20 \upmu \mathrm{m}$ the maximum is located at about Kn = 1 (fig. \ref{fig:tiefendruck} (center)).

This trend is consistent with earlier work by \citet{deBeule2015}. They find that the layer removed in their experiments at three pressures of 0.1, 1, and 10 mbar is largest at 1 mbar. So we clearly trace the same effect. More specific, we find that the maximum is situated at a Knudsen number of 1. 
As thermal creep is most efficient at Kn=1 \citep{Chambers2004} and does not work at large ambient pressure and less at lower ambient pressure \citep{Koester2017b},
this strongly supports the idea that thermal creep is the underlying mechanism for the particle motion.

\section{Caveats}

\subsection{Boundary effects}
The vessel holding the sample has a relatively small thickness. This is intended as it e.g. allows a faster cooling after one measurement before the next is started. However, it also means that the boundaries, i.e. the glass windows on the side have a significant influence on the temperature distribution. This cannot be avoided as we need a transparent wall that allows DWS at all and the particle columns need a stabilizing wall. With an order of magnitude of 1 W/(m K), the thermal conductivity of the walls is orders of magnitudes larger than the thermal conductivity of the sample. Therefore, the temperature of the window next to the relevant top few mm can be considered as being constant. So this is a heat source for the sample but overlaying as constant temperature it will not change the principle vertical stratification of the temperature differences and therefore, the position of the pressure maximum will only change slightly.
\subsection{Other motion}
The maximum in motion at around 10 mm (s. fig. \ref{fig:daten8mbar}) is also interesting but is not the focus of this work. We currently have no sophisticated model to explain these data. 
\subsection{Temperature dependence on pressure}
Fig. \ref{fig:tiefendruck} (bottom) shows that the overall temperature increases at low pressure but since it reaches a rather constant plateau, we do not consider this to influence our thermal creep measurements.

\section{Conclusion}

Thermal creep gas flow can significantly support the lifting of grains on Mars. However, the pressure variations are small and change on sub-mm spatial scales. Implementing arrays of commercial pressure sensors on sub-mm scale within a dust bed and resolutions on the Pa level without influencing the temperature profile if the dust bed is illuminated is currently virtually impossible. Therefore, the pressure gradient itself is hard to be measured directly.
With this lack of direct confirmation, even if the effect of thermal creep on larger scales is well known,  every indirect method can be an important verification. After all, thermal creep is not something encountered on Earth in natural settings and it does not come intuitively (we think).

Tracing the gas motion in and out of a dust bed was a milestone to attribute particle lifting to thermal creep gas flow \citep{debeule2014}. Observing the direct lifting of whole layers of grains in a work by \cite{deBeule2015} could pin down an active layer of $\sim 0.2$ mm within a dust bed, where we refer to activity as being under tension.

Here, this pressure maximum is visualized by subtle particle motions made visible by diffusing wave spectroscopy. The depth of about 2 mm is slightly larger than these earlier results but noting the different methodology and observing the same typical dependence of particle motion on ambient pressure, these findings are in agreement. Our results therefore provide one more confirmation of the effect of thermal creep under Martian conditions making thermal creep an ever more likely effect to influence the motion of grains on the Martian surface quite generally.

\section*{acknowledgements}
This project is supported by DLR Space Administration with funds provided by the Federal Ministry for Economic Affairs and Climate Action (BMWK) under grant numbers 50WM1943 and 50WM2049. This project also has received funding from the European
Union’s Horizon 2020 research and innovation
program under grant agreement No 101004052.
The manuscript gained significantly by the reviews of Frédéric Schmidt and an anonymous reviewer. 

\bibliography{Manuscript}

\begin{thebibliography}{}
\expandafter\ifx\csname natexlab\endcsname\relax\def\natexlab#1{#1}\fi
\providecommand{\url}[1]{\href{#1}{#1}}
\providecommand{\dodoi}[1]{doi:~\href{http://doi.org/#1}{\nolinkurl{#1}}}
\providecommand{\doeprint}[1]{\href{http://ascl.net/#1}{\nolinkurl{http://ascl.net/#1}}}
\providecommand{\doarXiv}[1]{\href{https://arxiv.org/abs/#1}{\nolinkurl{https://arxiv.org/abs/#1}}}

\bibitem[{Amon {et~al.}(2017)Amon, Mikhailovskaya, \& Crassous}]{Amon2017}
Amon, A., Mikhailovskaya, A., \& Crassous, J. 2017, Review of Scientific
  Instruments, 88, 051804, \dodoi{10.1063/1.4983048}

\bibitem[{{Balme} \& {Hagermann}(2006)}]{Balme2006}
{Balme}, M., \& {Hagermann}, A. 2006, \grl, 33, L19S01,
  \dodoi{10.1029/2006GL026819}

\bibitem[{{Bila} {et~al.}(2020){Bila}, {Wurm}, {Onyeagusi}, \&
  {Teiser}}]{Bila2020}
{Bila}, T., {Wurm}, G., {Onyeagusi}, F.~C., \& {Teiser}, J. 2020, \icarus, 339,
  113569, \dodoi{10.1016/j.icarus.2019.113569}

\bibitem[{Chambers(2004)}]{Chambers2004}
Chambers, A. 2004, Modern Vacuum Physics, Masters Series in Physics and
  Astronomy (CRC Press), 25--48

\bibitem[{{Chojnacki} {et~al.}(2019){Chojnacki}, {Banks}, {Fenton}, \&
  {Urso}}]{Chojnacki2019}
{Chojnacki}, M., {Banks}, M.~E., {Fenton}, L.~K., \& {Urso}, A.~C. 2019,
  Geology, 47, 427, \dodoi{10.1130/G45793.1}

\bibitem[{Crassous(2007)}]{Crassous2007}
Crassous, J. 2007, The European Physical Journal E, 23, 145,
  \dodoi{10.1140/epje/i2006-10079-y}

\bibitem[{{de Beule} {et~al.}(2015){de Beule}, {Wurm}, {Kelling}, {Koester}, \&
  {Kocifaj}}]{deBeule2015}
{de Beule}, C., {Wurm}, G., {Kelling}, T., {Koester}, M., \& {Kocifaj}, M.
  2015, \icarus, 260, 23, \dodoi{10.1016/j.icarus.2015.06.002}

\bibitem[{{de Beule} {et~al.}(2014){de Beule}, {Wurm}, {Kelling}, {K{\"u}pper},
  {Jankowski}, \& {Teiser}}]{debeule2014}
{de Beule}, C., {Wurm}, G., {Kelling}, T., {et~al.} 2014, Nature Physics, 10,
  17, \dodoi{10.1038/nphys2821}

\bibitem[{Erpelding {et~al.}(2008)Erpelding, Amon, \& Crassous}]{Erpelding2008}
Erpelding, M., Amon, A., \& Crassous, J. 2008, Physical Review E, 78,
  \dodoi{10.1103/physreve.78.046104}

\bibitem[{Erpelding {et~al.}(2010)Erpelding, Amon, \&
  Crassous}]{Erpelding_2010}
---. 2010, {EPL} (Europhysics Letters), 91, 18002,
  \dodoi{10.1209/0295-5075/91/18002}

\bibitem[{Esposito {et~al.}(2016)Esposito, Molinaro, Popa, Molfese, Cozzolino,
  Marty, Taj-Eddine, Di~Achille, Franzese, Silvestro, \& Ori}]{Esposito2016}
Esposito, F., Molinaro, R., Popa, C.~I., {et~al.} 2016, Geophysical Research
  Letters, 43, 5501, \dodoi{10.1002/2016GL068463}

\bibitem[{{Fenton}(2020)}]{Fenton2020}
{Fenton}, L.~K. 2020, \icarus, 352, 114018,
  \dodoi{10.1016/j.icarus.2020.114018}

\bibitem[{{Greeley} {et~al.}(1980){Greeley}, {Leach}, {White}, {Iversen}, \&
  {Pollack}}]{Greeley1980}
{Greeley}, R., {Leach}, R., {White}, B., {Iversen}, J., \& {Pollack}, J.~B.
  1980, \grl, 7, 121, \dodoi{10.1029/GL007i002p00121}

\bibitem[{{Heyer} {et~al.}(2020){Heyer}, {Raack}, {Hiesinger}, \&
  {Jaumann}}]{Heyer2020}
{Heyer}, T., {Raack}, J., {Hiesinger}, H., \& {Jaumann}, R. 2020, \icarus, 351,
  113951, \dodoi{10.1016/j.icarus.2020.113951}

\bibitem[{{Kelling} \& {Wurm}(2009)}]{Kelling2009}
{Kelling}, T., \& {Wurm}, G. 2009, \prl, 103, 215502,
  \dodoi{10.1103/PhysRevLett.103.215502}

\bibitem[{{Kelling} {et~al.}(2011){Kelling}, {Wurm}, {Kocifaj}, {Kla{\v{c}}ka},
  \& {Reiss}}]{Kelling2011}
{Kelling}, T., {Wurm}, G., {Kocifaj}, M., {Kla{\v{c}}ka}, J., \& {Reiss}, D.
  2011, \icarus, 212, 935, \dodoi{10.1016/j.icarus.2011.01.010}

\bibitem[{{Knudsen}(1910)}]{Knudsen1910}
{Knudsen}, M. 1910, Annalen der Physik, 336, 633,
  \dodoi{10.1002/andp.19103360310}

\bibitem[{{Kocifaj} {et~al.}(2011){Kocifaj}, {Kla{\v{c}}ka}, {Kelling}, \&
  {Wurm}}]{Kocifaj2011}
{Kocifaj}, M., {Kla{\v{c}}ka}, J., {Kelling}, T., \& {Wurm}, G. 2011, \icarus,
  211, 832, \dodoi{10.1016/j.icarus.2010.10.006}

\bibitem[{{Kocifaj} {et~al.}(2010){Kocifaj}, {Kla{\v{c}}ka}, {Wurm}, {Kelling},
  \& {Koh{\'u}t}}]{Kocifaj2010}
{Kocifaj}, M., {Kla{\v{c}}ka}, J., {Wurm}, G., {Kelling}, T., \& {Koh{\'u}t},
  I. 2010, \mnras, 404, 1512, \dodoi{10.1111/j.1365-2966.2010.16370.x}

\bibitem[{{Koester} {et~al.}(2017){Koester}, {Kelling}, {Teiser}, \&
  {Wurm}}]{Koester2017b}
{Koester}, M., {Kelling}, T., {Teiser}, J., \& {Wurm}, G. 2017, \apss, 362,
  171, \dodoi{10.1007/s10509-017-3154-4}

\bibitem[{{Koester} \& {Wurm}(2017)}]{Koester2017}
{Koester}, M., \& {Wurm}, G. 2017, \planss, 145, 9,
  \dodoi{10.1016/j.pss.2017.07.005}

\bibitem[{Kok(2010)}]{Kok2010b}
Kok, J.~F. 2010, Phys. Rev. Lett., 104, 074502,
  \dodoi{10.1103/PhysRevLett.104.074502}

\bibitem[{{Kok} {et~al.}(2012){Kok}, {Parteli}, {Michaels}, \&
  {Karam}}]{Kok2012}
{Kok}, J.~F., {Parteli}, E. J.~R., {Michaels}, T.~I., \& {Karam}, D.~B. 2012,
  Reports on Progress in Physics, 75, 106901,
  \dodoi{10.1088/0034-4885/75/10/106901}

\bibitem[{{Kruss} {et~al.}(2021){Kruss}, {Salzmann}, {Parteli}, {Jungmann},
  {Teiser}, {Sch{\"o}nau}, \& {Wurm}}]{Kruss2021}
{Kruss}, M., {Salzmann}, T., {Parteli}, E., {et~al.} 2021, Planetary Science
  Journal, 2, 238, \dodoi{10.3847/PSJ/ac38a4}

\bibitem[{{Kuepper} \& {Wurm}(2016)}]{Kuepper2016}
{Kuepper}, M., \& {Wurm}, G. 2016, \icarus, 274, 249,
  \dodoi{10.1016/j.icarus.2016.02.049}

\bibitem[{{K{\"u}pper} \& {Wurm}(2015)}]{Kuepper2015}
{K{\"u}pper}, M., \& {Wurm}, G. 2015, Journal of Geophysical Research
  (Planets), 120, 1346, \dodoi{10.1002/2015JE004848}

\bibitem[{{Lorenz} {et~al.}(2021){Lorenz}, {Lemmon}, \& {Maki}}]{Lorenz2021}
{Lorenz}, R.~D., {Lemmon}, M.~T., \& {Maki}, J. 2021, \icarus, 364, 114468,
  \dodoi{10.1016/j.icarus.2021.114468}

\bibitem[{{Merrison} {et~al.}(2007){Merrison}, {Gunnlaugsson}, {N{\o}rnberg},
  {Jensen}, \& {Rasmussen}}]{Merrison2007}
{Merrison}, J.~P., {Gunnlaugsson}, H.~P., {N{\o}rnberg}, P., {Jensen}, A.~E.,
  \& {Rasmussen}, K.~R. 2007, \icarus, 191, 568,
  \dodoi{10.1016/j.icarus.2007.04.035}

\bibitem[{{Musiolik} {et~al.}(2018){Musiolik}, {Kruss}, {Demirci}, {Schrinski},
  {Teiser}, {Daerden}, {Smith}, {Neary}, \& {Wurm}}]{Musiolik2018}
{Musiolik}, G., {Kruss}, M., {Demirci}, T., {et~al.} 2018, \icarus, 306, 25,
  \dodoi{10.1016/j.icarus.2018.01.007}

\bibitem[{{Neakrase} {et~al.}(2016){Neakrase}, {Balme}, {Esposito}, {Kelling},
  {Klose}, {Kok}, {Marticorena}, {Merrison}, {Patel}, \& {Wurm}}]{Neakrase2016}
{Neakrase}, L.~D.~V., {Balme}, M.~R., {Esposito}, F., {et~al.} 2016, \ssr, 203,
  347, \dodoi{10.1007/s11214-016-0296-6}

\bibitem[{{Raack} {et~al.}(2017){Raack}, {Conway}, {Herny}, {Balme}, {Carpy},
  \& {Patel}}]{Raack2017}
{Raack}, J., {Conway}, S.~J., {Herny}, C., {et~al.} 2017, Nature
  Communications, 8, 1151, \dodoi{10.1038/s41467-017-01213-z}

\bibitem[{{Rasmussen} {et~al.}(2015){Rasmussen}, {Valance}, \&
  {Merrison}}]{Rasmussen2015}
{Rasmussen}, K.~R., {Valance}, A., \& {Merrison}, J. 2015, Geomorphology, 244,
  74, \dodoi{10.1016/j.geomorph.2015.03.041}

\bibitem[{{Schmidt} {et~al.}(2017){Schmidt}, {Andrieu}, {Costard}, {Kocifaj},
  \& {Meresescu}}]{Schmidt2017}
{Schmidt}, F., {Andrieu}, F., {Costard}, F., {Kocifaj}, M., \& {Meresescu},
  A.~G. 2017, Nature Geoscience, 10, 270, \dodoi{10.1038/ngeo2917}

\bibitem[{Steinpilz {et~al.}(2017)Steinpilz, Teiser, Koester, Schywek, \&
  Wurm}]{Steinpilz2017}
Steinpilz, T., Teiser, J., Koester, M., Schywek, M., \& Wurm, G. 2017, MicST,
  29, 325, \dodoi{10.1007/s12217-017-9550-0}

\bibitem[{{Swann} {et~al.}(2020){Swann}, {Sherman}, \& {Ewing}}]{Swann2020}
{Swann}, C., {Sherman}, D.~J., \& {Ewing}, R.~C. 2020, \grl, 47, e84484,
  \dodoi{10.1029/2019GL084484}

\bibitem[{{Toigo} {et~al.}(2018){Toigo}, {Richardson}, {Wang}, {Guzewich}, \&
  {Newman}}]{Toigo2018}
{Toigo}, A.~D., {Richardson}, M.~I., {Wang}, H., {Guzewich}, S.~D., \&
  {Newman}, C.~E. 2018, \icarus, 302, 514, \dodoi{10.1016/j.icarus.2017.11.032}

\bibitem[{{Vi{\'u}dez-Moreiras} {et~al.}(2020){Vi{\'u}dez-Moreiras}, {Newman},
  {Forget}, {Lemmon}, {Banfield}, {Spiga}, {Lepinette}, {Rodriguez-Manfredi},
  {G{\'o}mez-Elvira}, {Pla-Garc{\'\i}a}, {Muller}, \& {Grott}}]{Moreiras2020}
{Vi{\'u}dez-Moreiras}, D., {Newman}, C.~E., {Forget}, F., {et~al.} 2020,
  Journal of Geophysical Research (Planets), 125, e06493,
  \dodoi{10.1029/2020JE006493}

\bibitem[{Weitz \& Pine(1993)}]{brown1993}
Weitz, D.~A., \& Pine, D.~J. 1993, in Monographs on the physics and chemistry
  of materials, Vol.~49, Dynamic Light Scattering: The Method and Some
  Applications, ed. W.~Brown (Clarendon Press), 652 -- 720

\bibitem[{{Wurm} \& {Krauss}(2006)}]{Wurm2006}
{Wurm}, G., \& {Krauss}, O. 2006, \prl, 96, 134301,
  \dodoi{10.1103/PhysRevLett.96.134301}

\bibitem[{{Wurm} {et~al.}(2008){Wurm}, {Teiser}, \& {Reiss}}]{Wurm2008}
{Wurm}, G., {Teiser}, J., \& {Reiss}, D. 2008, \grl, 35, L10201,
  \dodoi{10.1029/2008GL033799}

\end{thebibliography}

\end{document}